\documentclass[12pt]{article}
\usepackage{amsfonts}
\usepackage{amsmath}
\usepackage{amssymb}
\usepackage{array}
\usepackage{bigints}
\usepackage{booktabs}
\usepackage[nosort]{cite}
\usepackage{color}
\usepackage{dsfont}
\usepackage{float}
\usepackage{framed}
\usepackage{graphicx}
\usepackage{indentfirst}
\usepackage{mathrsfs}
\usepackage{multirow}
\usepackage{setspace}
\usepackage{subdepth}
\usepackage{subfig}
\usepackage{titlesec}
\usepackage[dotinlabels]{titletoc}
\usepackage{wrapfig}
\usepackage[all]{xy}
\usepackage{young}
\usepackage[vcentermath]{youngtab}
\usepackage{relsize}

\usepackage{hyperref}
\hypersetup{colorlinks=true}
\hypersetup{linkcolor=black}
\hypersetup{citecolor=black}
\hypersetup{urlcolor=black}

\numberwithin{equation}{section}

\usepackage[left=2.5cm,right=2.5cm,top=2.5cm,bottom=3cm]{geometry}
\titleformat{\section}{\normalfont\bfseries}{\thesection.}{4pt}{}
\titlespacing{\section}{0pt}{20pt}{6pt}

\titleformat{\subsection}{\normalfont\itshape}{\thesubsection.}{4pt}{}
\titlespacing{\subsection}{0pt}{15pt}{6pt}

\titleformat{\subsubsection}{\normalfont}{\thesubsubsection.}{4pt}{}
\titlespacing{\subsubsection}{0pt}{15pt}{6pt}

\def\ie{\begin{equation}\begin{aligned}}
\def\fe{\end{aligned}\end{equation}}

\newcommand\VRule[1][\arrayrulewidth]{\vrule width #1}

\def\hat{\widehat}

\def\half{{1 \over 2}}
\def\d{\partial}

\def\ep{\varepsilon}
\def\1{{\mathds 1}}

\DeclareMathOperator{\tr}{tr}
\DeclareMathOperator{\Tr}{\mathrm{Tr}}

\newcommand{\Z}{{\mathbb Z}}

\def\SL{{\mathscr L}}

\def\CE{{\mathcal E}}

\def\CI{{\mathcal I}}

\def\CN{{\mathcal N}}
\def\CO{{\mathcal O}}

\def\CT{{\mathcal T}}

\DeclareFontShape{OT1}{cmr}{mx}{n}%
    {<->cmr10}{}
\newcommand{\mytitlefont}{\fontseries{mx}\selectfont}
\DeclareMathAlphabet{\titlemath}{OT1}{cmr}{mx}{n}

\begin{document}


\begin{titlepage}

\begin{center}

~\\[2cm]

{\fontsize{22pt}{0pt} \mytitlefont Anomalies, Renormalization Group Flows, and \\[4pt]
the~$a$-Theorem in Six-Dimensional $(1,0)$ Theories}

~\\[0.5cm]

Clay C\'{o}rdova,$^1$ Thomas T.~Dumitrescu,$^2$ and Kenneth Intriligator\,$^3$

~\\[0.1cm]

$^1$~{\it Society of Fellows, Harvard University, Cambridge, MA 02138, USA}

$^2$~{\it Department of Physics, Harvard University, Cambridge, MA 02138, USA}

$^3$ {\it Department of Physics, University of California, San Diego, La Jolla, CA 92093, USA}

~\\[0.8cm]

\end{center}

\noindent We establish a linear relation between the~$a$-type Weyl anomaly and the 't Hooft anomaly coefficients for the~$R$-symmetry and gravitational anomalies in six-dimensional $(1,0)$ superconformal field theories. For RG flows onto the tensor branch, where conformal symmetry is spontaneously broken, supersymmetry relates the anomaly mismatch~$\Delta a$ to the square of a four-derivative interaction for the dilaton. This establishes the~$a$-theorem for all such flows. The four-derivative dilaton interaction is in turn related to the Green-Schwarz-like terms that are needed to match the 't Hooft anomalies on the tensor branch, thus fixing their relation to~$\Delta a$. We use our formula to obtain exact expressions for the~$a$-anomaly of~$N$ small $E_8$ instantons, as well as~$N$ M5-branes probing an orbifold singularity, and verify the~$a$-theorem for RG flows onto their Higgs branches. We also discuss aspects of supersymmetric RG flows that terminate in scale but not conformally invariant theories with massless gauge fields.  
\vfill

\begin{flushleft}
June 2015
\end{flushleft}

\end{titlepage}


\tableofcontents

\section{Introduction}
\label{sec:intro}

A basic set of observables that exists for all conformal field theories (CFTs) in even spacetime dimensions is furnished by the Weyl anomalies, which can be defined through the anomalous trace of the stress tensor~$T_{\mu\nu}$ in the presence of a background metric~\cite{Deser:1976yx,Duff:1977ay,Fradkin:1983tg,Deser:1993yx},
\begin{equation}
\langle T_\mu ^\mu \rangle \sim a E_d+\sum _i c_i I_i ~,\label{Tanomaly}
\end{equation}
up to scheme-dependent terms. Here~$E_d$ is the~$d$-dimensional Euler density, and the~$I_i$ are local Weyl invariants of weight~$d$, whose number depends on the spacetime dimension. The dimensionless anomaly coefficients~$a, c_i$ also appear in flat-space correlation functions of~$T_{\mu\nu}$ at separated points. 

The~$a$-anomaly plays an important role in the study of renormalization group (RG) flows. In two and four dimensions, it was shown that all unitary RG flows between CFTs in the UV and in the IR satisfy the~$a$-theorem, which states that\footnote{~Note that the two-dimensional~$a$-anomaly is usually denoted by~$c$, since it coincides with the Virasoro central charge. There are no~$c$-type anomalies in two dimensions.}
\begin{equation}
\Delta a =  a_{\text{UV}}-a_{\text{IR}}>0~.\label{atheorem}
\end{equation}
The two-dimensional version was established in~\cite{Zamolodchikov:1986gt}. The four-dimensional~$a$-theorem was conjectured in~\cite{Cardy:1988cwa} and further analyzed in~\cite{Osborn:1989td,Jack:1990eb}. A proof was presented in~\cite{Komargodski:2011vj,Komargodski:2011xv}. It utilizes the fact that the anomaly matching conditions for the~$a$-anomaly discussed in~\cite{Schwimmer:2010za} lead to a special Wess-Zumino-like interaction term in the effective action for a (dynamical or background) dilaton field, which is interpreted as a Nambu-Goldstone (NG) boson for the spontaneous breaking of conformal symmetry. Another property of the~$a$-anomaly in two and four dimensions that is closely related to, but independent of~\eqref{atheorem} is that it is non-negative,
\begin{equation}
a \geq 0~,\label{apositivity}
\end{equation}
and vanishes if and only if the theory has no local degrees of freedom~\cite{Zamolodchikov:1986gt,Hofman:2008ar}.  

Prior to~\cite{Komargodski:2011vj,Komargodski:2011xv}, some of the strongest evidence for the four-dimensional~$a$-theorem came from supersymmetric RG flows, see for instance~\cite{Anselmi:1997am,Intriligator:2003jj,Kutasov:2003iy,Intriligator:2003mi,Erkal:2010sh}. For~$\CN=1$ superconformal field theories (SCFTs) in~$d =4$ dimensions, it was shown in~\cite{Anselmi:1997am} that the~$a$-anomaly is linearly related to the 't Hooft anomalies for the~$U(1)_R$ symmetry,
\begin{equation}
a =\frac{3}{32}(3 k_{RRR}-k_R)~. \label{a4d}
\end{equation}
Here~$k_{RRR}$ and~$k_R$ are the~$\Tr U(1)_R^3$ and~$\Tr U(1)_R$ 't Hooft anomalies, which appear in the anomaly polynomial,\footnote{~See section~\ref{sec:poly} for a brief review of anomaly polynomials.}
\begin{equation}
{\cal I}_6 = {1 \over 3 !} \left(k_{RRR} \, c_1(R)^3 + k_R \,c_1(R)p_1(T)\right)~.
\end{equation}
The formula~\eqref{a4d} makes it possible to determine the~$a$-anomaly in a large number of strongly interacting examples, by relying on the calculability of 't Hooft anomalies.

Given the two- and four-dimensional results summarized above, it is natural to anticipate similar statements in six dimensions, but the proof of a six-dimensional~$a$-theorem remains an open problem. (See~\cite{Myers:2010tj} and references therein for evidence from holography.) Following~\cite{Komargodski:2011vj,Komargodski:2011xv}, the constraints of conformal symmetry on dilaton self-interactions in six dimensions, and their relation to the~$a$-anomaly, were analyzed in~\cite{Maxfield:2012aw, Elvang:2012st}, where it was pointed out that~$\Delta a$ can in general receive contributions of either sign, even in a unitary theory (see also~\cite{Grinstein:2014xba, Grinstein:2015ina}). The ability to test the six-dimensional~$a$-theorem is limited by the fact that the~$a$-anomaly has only been computed for a handful of interacting CFTs, and that examples of controlled RG flows between such theories are scarce. 

All known examples of interacting CFTs in six dimensions are supersymmetric and arise from decoupling limits of string constructions. Nevertheless, they are believed to be quantum field theories~\cite{Seiberg:1996vs,Seiberg:1996qx}. The most well-studied such theories are the maximally supersymmetric~$(2,0)$ SCFTs constructed in~\cite{Witten:1995zh, Strominger:1995ac, Witten:1995em}, which are labeled by an ADE Lie algebra~$\frak g$. The~$A$-type theories arise on the worldvolume of parallel M5-branes in~M-theory~\cite{Strominger:1995ac}.  Their 't Hooft anomalies have been computed in~\cite{Duff:1995wd,Witten:1996hc,Freed:1998tg,Harvey:1998bx,Intriligator:2000eq,Yi:2001bz,Ohmori:2014kda}. Some of the~$(2,0)$ theories come in infinite families that admit a large-$N$ limit with weakly coupled holographic duals, and these have been used to show that the~$a$-anomaly scales like~$N^3$ at leading order in the~$1 \over N$ expansion~\cite{Henningson:1998gx}. Some subleading corrections were discussed in~\cite{Tseytlin:2000sf,Beccaria:2014qea}. The~$(2,0)$ theories have a moduli space of vacua on which conformal symmetry is spontaneously broken, and (partially) moving onto this moduli space induces a non-trivial RG flow.\footnote{~One might distinguish RG flows onto the moduli space, where conformal invariance is broken spontaneously, from those associated with explicit breaking, which are triggered by adding a relevant operator.  In either case, there is a flow along which one integrates out massive degrees of freedom. }  The constraints of maximal supersymmetry on such flows were systematically analyzed in~\cite{Cordova:2015vwa}, following~\cite{Maxfield:2012aw}, which lead to an~exact calculation of the~$a$-anomaly for all~$(2,0)$ theories and a proof of the $a$-theorem for all RG flows that preserve $(2,0)$ supersymmetry.  

A large class of interacting six-dimensional SCFTs with~$(1,0)$ supersymmetry have been constructed in string theory, starting with the work of \cite{Seiberg:1996vs, Seiberg:1996qx}. Further examples were studied using brane constructions~\cite{Ganor:1996mu, Blum:1997mm,Brunner:1997gk,Brunner:1997gf,Hanany:1997gh}. Recently, vast landscapes of~$(1,0)$ theories have been systematically constructed in F-theory \cite{Heckman:2013pva, DelZotto:2014hpa, Heckman:2015bfa}, and a detailed analysis of holographic theories with~$(1,0)$ supersymmetry was carried out in~\cite{Apruzzi:2013yva, Gaiotto:2014lca}. Large classes of RG flows in these examples were studied in~\cite{Gaiotto:2014lca,Heckman:2015ola}. All of these flows are induced by moving onto a moduli space of vacua. This is a general feature of all~$(1,0)$ SCFTs: using superconformal representation theory, it can be shown that such theories do not contain relevant or marginal operators that can be used to deform the theory while preserving supersymmetry~\cite{Cordova:2016xhm,Cordova:2016emh} (see also~\cite{Louis:2015mka}), and hence all supersymmetric RG flows are necessarily moduli-space flows. 

Since all known six-dimensional interacting CFTs are supersymmetric, it has been a longstanding expectation (see for instance~\cite{Harvey:1998bx}) that supersymmetry should make it possible to compute the~$a$-anomaly in these theories by relating it to their 't Hooft anomalies, in analogy with the known relations~\eqref{a4d} in four dimensions. Such a relation is expected to follow from an anomalous stress-tensor supermultiplet, which embeds the anomalous trace of the stress tensor in~\eqref{Tanomaly} into rigid background supergravity.  These anomaly multiplets are currently under investigation~\cite{ctkii}. Another ambitious line of attack would be to directly supersymmetrize the four-point functions of stress-tensors and~$R$-currents, in analogy with the results of~\cite{Osborn:1998qu} on stress-tensor three-point functions in four dimensions.  

In this paper, we assume the existence of a universal linear relation, valid for all~$(1,0)$ SCFTs, that relates the~$a$-anomaly to the 't Hooft anomalies. Given this assumption, we then derive the precise formula by combining the constraints of supersymmetry with those from anomaly matching for the Weyl and 't Hooft anomalies. We find that 
\begin{equation}
a= \frac{16}{7}\left(\alpha -  \beta +  \gamma \right)+ \frac{6}{7} \delta~.\label{ais}
\end{equation}
Here, and throughout the paper, we use a normalization of the~$a$-anomaly, in which a free~$(2,0)$ tensor multiplet has~$a = 1$. The constants~$\alpha, \beta, \gamma, \delta$ are 't Hooft anomaly coefficients for the~$SU(2)_R$ symmetry and gravitational anomalies of the theory, which enter the anomaly polynomial as follows,\footnote{~We follow the conventions of~\cite{Ohmori:2014kda,Intriligator:2014eaa} for anomaly polynomials and characteristic classes.}\begin{equation}
\CI_8 = {1 \over 4!} \Big(\alpha c_2^2(R) + \beta c_2(R) p_1(T) + \gamma p_1^2(T) + \delta p_2(T)\Big)~.\label{anomalyI}
\end{equation}
In Table~\ref{known}, we summarize the values of the~$a$-anomaly and the 't Hooft anomaly coefficients in~\eqref{anomalyI} for all~$(1,0)$ free fields, which are known from~\cite{Fradkin:1983tg, AlvarezGaume:1983ig, Bastianelli:2000hi}, as well as for all~$(2,0)$ theories. The negative value of~$a$ for a free vector multiplet, which is not a CFT, is obtained by naively applying~\eqref{ais}. Its meaning will be discussed in detail below.

\begin{table}[h]
\centering
\begin{tabular}{!{\VRule[1pt]}c!{\VRule[1pt]}c!{\VRule[1pt]}c!{\VRule[1pt]}c!{\VRule[1pt]} c!{\VRule[1pt]}c!{\VRule[1pt]}}
\specialrule{1.2pt}{0pt}{0pt}
{\bf Theory} & $\bf \alpha$ &  $\bf \beta$ & $\bf \gamma$ & $\bf \delta$ & $a$ \rule{0pt}{2.6ex}\rule[-1.4ex]{0pt}{0pt} \\
\specialrule{1.2pt}{0pt}{0pt}
\multirow{2}{*}{ Hypermultiplet}& \multirow{2}{*}{$0$} &  \multirow{2}{*}{$0$} & \multirow{2}{*}{$7 \over 240$} & \multirow{2}{*}{$-{1 \over 60}$} & \multirow{2}{*}{$11 \over 210$}\\
 &  & & & & \\
\hline
\multirow{2}{*}{Tensor multiplet}& \multirow{2}{*}{$1$} &  \multirow{2}{*}{$\half$} & \multirow{2}{*}{$23 \over 240$} & \multirow{2}{*}{$-{29 \over 60}$} & \multirow{2}{*}{$199 \over 210$}\\
 &  & & & & \\
 \hline
 \multirow{2}{*}{Vector multiplet}& \multirow{2}{*}{$-1$} &  \multirow{2}{*}{$- \half $} & \multirow{2}{*}{$-{7 \over 240}$} & \multirow{2}{*}{$1 \over 60$}& \multirow{2}{*}{``\,$- {251 \over 210}$\, ''}\\
 &  & & & & \\
\hline
 \multirow{2}{*}{$(2,0)$ Theory with algebra $\frak{g}$}& \multirow{2}{*}{$h^\vee _{\frak{g}}d_{\frak{g}}+r_{\frak{g}}$} &  \multirow{2}{*}{$\half r_{\frak{g}} $} & \multirow{2}{*}{$\frac{1}{8}r_{\frak{g}}$} & \multirow{2}{*}{$-\half r_{\frak{g}}$}& \multirow{2}{*}{$\frac{16}{7}h_{\frak{g}}^\vee d_{\frak{g}}+r_{\frak{g}}$}\\
 &  & & & & \\
\specialrule{1.2pt}{0pt}{0pt}
 \end{tabular}
\caption{'t Hooft and $a$-anomalies for known examples. (See section~\ref{sec:vectors} for a detailed discussion of the vector multiplet.)}
\label{known}
\end{table}

In section~\ref{sec:anomaly}, we begin by reviewing aspects of anomalies and anomaly matching in six-dimensional~$(1,0)$ theories. In particular, we review the Green-Schwarz (GS) like anomaly matching mechanism for 't Hooft anomalies on the tensor branch described in~\cite{Intriligator:2000eq, Ohmori:2014kda,Intriligator:2014eaa}. For the case of a single tensor multiplet, a GS term of the form\footnote{~The factor of~$i$ is due to the fact that we are working in Euclidean signature.} 
\begin{equation} 
 -i B\wedge X_4 \subset \SL~, \qquad X_4 \sim x \, c_2(R) + y \, p_1(T)~,\label{BXfour}
\end{equation}
contributes a perfect square to the anomaly polynomial,
\begin{equation}
\Delta \CI  _8 \sim  X_4\wedge X_{4}~.\label{DeltaIsq}
\end{equation}
As a result, the 't Hooft anomaly coefficients in~\eqref{anomalyI} satisfy
\begin{equation}
\Delta \alpha \sim x^2~, \quad \Delta \beta \sim 2xy~, \quad \Delta \gamma \sim y^2~, \quad \Delta \delta = 0~. \label{rk1square}
\end{equation}

In section~\ref{sec:anomaly} we also review the results of~\cite{Maxfield:2012aw, Elvang:2012st,Elvang:2012yc} on the constraints of conformal symmetry on the dilaton effective Lagrangian in six dimensions. In particular, we recall that the mismatch~$\Delta a$ in the~$a$-anomaly arises as the coefficient of a particular six-derivative interaction term for the dilaton. 

In section~\ref{sec:athm} we analyze the constraints of~$(1,0)$ supersymmetry on the dilaton effective action for tensor branch flows. While the coefficient~$b$ that controls the four-derivative interactions of the dilaton is unconstrained, the six-derivative terms satisfy a non-renormalization theorem that leads to a quadratic relation of the form
\begin{equation}
\Delta a \sim b^{2}~, \label{adiff}
\end{equation}
with a positive, model-independent proportionality factor that will be discussed in section~\ref{sec:athm}. An identical quadratic relation was known to hold on the moduli space of~$(2,0)$ SCFTs~\cite{Maxfield:2012aw,Elvang:2012st, Cordova:2015vwa, Chen:2015hpa}, but it is also valid in theories with~$(1,0)$ supersymmetry. (See also the recent discussion in~\cite{Chen:2015hpa}.) The relation~\eqref{adiff} immediately implies the~$a$-theorem for RG flows onto the tensor branch, as was emphasized for~$(2,0)$ theories in~\cite{Cordova:2015vwa}. 

In section~\ref{sec:relation}, we derive the anomaly relation~\eqref{ais}. As in section~\ref{sec:athm}, we consider RG flows onto the tensor branch and show that the changes in the anomaly coefficients along such flows must satisfy
\begin{equation}\label{deltaaintro}
\Delta a = {16 \over 7} \left(\Delta \alpha - \Delta \beta + \Delta \gamma\right)~.
\end{equation}
In order to establish this relation, we couple the theory to background conformal supergravity fields and use the results of~\cite{Bergshoeff:1986wc} on higher-curvature terms in this supergravity theory. This reveals a universal linear relation between the GS term~\eqref{BXfour} and the coefficient~$b$ that controls the four-derivative dilaton interactions. Since the former determines~$\Delta \alpha, \Delta \beta, \Delta \gamma$ via~\eqref{DeltaIsq} and the latter is quadratically related to~$\Delta a$ via~\eqref{adiff}, we obtain~\eqref{deltaaintro}. This only leaves the coefficient of~$\delta$ in~\eqref{ais} undetermined, which can be fixed by examining the anomalies of a free hypermultiplet. The known values of the 't Hooft and $a$-anomalies for the free tensor multiplet and the~$(2,0)$ theories in Table~\ref{known} constitute non-trivial consistency checks.\footnote{~The relation between $a$ and the 't Hooft anomalies can in principle be determined by fitting a linear formula using sufficiently many reliable examples. The free theories and the~$(2,0)$ theories can be used to fix~\eqref{ais} up to one undetermined coefficient.}

In section~\ref{sec:E8}, we apply our results to compute the~$a$-anomaly for the theory~$\mathcal{E}_{N}$ of $N$ small $E_8$ instantons~\cite{Ganor:1996mu}, whose anomaly polynomial was computed in~\cite{Ohmori:2014pca},\footnote{~We define both the small instanton theory and the orbifold examples to include their free center of mass modes.}
\begin{equation}
a(\mathcal{E}_{N})=\frac{64}{7}N^{3}+\frac{144}{7}N^{2}+\frac{99}{7}N~.   \label{aE8is}
\end{equation}
Similarly, the theory~$\mathcal{T}_{N,\Gamma}$ of $N$ M5-branes probing a~$\mathbb{C}^2/\Gamma$ orbifold singularity, whose 't Hooft anomalies were obtained in~\cite{Ohmori:2014pca}, has the following~$a$-anomaly,
\begin{equation}
 a(\mathcal{T}_{N,\Gamma})=\frac{16}{7}N^{3}|\Gamma|^{2}-\frac{24}{7}N|\Gamma|(r_{\Gamma}+1)+\frac{15}{7}N+\frac{251}{210}d_{\Gamma}~. \label{aorbis}
\end{equation}
This example is discussed in section~\ref{sec:orbifold}. We also verify that~$\Delta a > 0$ for some Higgs branch flows in these theories, which does not automatically  follow from our general arguments. 

In section~\ref{sec:vectors}, we discuss subtleties in the statement and proof of an~$a$-theorem for RG flows that terminate on tensor branches with vector multiplets.  Unlike in four dimensions, where free gauge fields constitute an ordinary CFT, in six dimensions the free vector field is scale invariant, but not conformally invariant, i.e. it is an SFT.  (See for instance~\cite{ElShowk:2011gz}.) Such theories possess a well-defined stress tensor~$T_{\mu\nu}$, but its trace~$T^\mu_\mu \sim \Tr (f^2)$ does not vanish. (Here~$f$ is the field strength).  String constructions of six-dimensional field theories suggest that RG flows from CFTs to free SFTs abound.  To formulate an~$a$-theorem for such flows, we must extend the definition of the~$a$-anomaly to these theories, and we do so by insisting that~\eqref{ais} holds.  With this definition, an $a$-theorem for tensor branch flows continues to hold, but~$a$ may no longer be positive (see for instance the vector multiplet in Table \ref{known}). Nevertheless, the~$a$-anomaly for the UV conformal field theory turns out to be positive in the examples we consider. 

In appendix~\ref{app}, we review the GS mechanism for chiral scalars in two dimensions, to supplement the six-dimensional discussion in section~\ref{sec:moduli}.

\section{Anomaly Matching in Six Dimensions}
\label{sec:anomaly}

In this section we review some necessary background material about anomalies. In particular, we explain how the~$SU(2)_R$ and gravitational 't Hooft anomalies, as well as the~$a$-type Weyl anomaly, are matched on the tensor and Higgs branches of~$(1,0)$ SCFTs. 

\subsection{Anomaly Polynomials}
\label{sec:poly}

Throughout this paper we will work in Euclidean signature. In even spacetime dimensions~$d=2n$, conventional local anomalies are encoded by a~$(d+2)$-form~$\CI_{d+2}$ residing in~$d+2$ dimensions, which is a polynomial in the Chern and Pontryagin classes of dynamical or background gauge and gravity fields. (We follow the conventions of~\cite{Ohmori:2014kda,Intriligator:2014eaa} for anomaly polynomials and characteristic classes.) For this reason~$\CI_{d+2}$ is also known as the anomaly polynomial. Under a gauge transformation or diffeomorphism~$\delta$, the anomalous variation of the Euclidean effective Lagrangian~$\SL$, which enters the path integral via~$e^{-\int \SL}$, is given by
\begin{equation}
\delta \SL=2\pi i \, \CI_d~.
\end{equation}
Here~$\CI_d$ is a differential~$d$-form polynomial in the gauge and gravity fields, which can be obtained from~$\CI_{d+2}$ via the descent procedure~\cite{AlvarezGaume:1983ig,AlvarezGaume:1984dr,Bardeen:1984pm},
\begin{equation}\label{descent}
\CI _{d+2}=d\CI _{d+1}~, \qquad \delta \CI _{d+1}=d\CI_{d}~,
\end{equation}
where a subscript~$p$ indicates a differential~$p$-form. 

Broadly speaking, we can group the terms in the anomaly polynomial as follows,
\begin{equation}\label{Iterms}
\CI_{d+2} =\CI^{\text{gauge}}_{d+2}+\CI ^{\text{gravity}}_{d+2}+\CI^{\text{mixed}}_{d+2}~.
\end{equation}
The terms in~$\CI^{\text{gauge}}_{d+2}$ are monomials in the Chern classes~$c_k(f)$ for a dynamical or background gauge field~$f$; they are~$2k$-forms, as well as Casimir invariants of the gauge group. If the gauge field is dynamical, the theory is only consistent if all of its anomalies vanish or can be cancelled.  The anomaly corresponding to~$c_{d+2}(f)$ is irreducible, i.e.~it cannot be cancelled, and hence it must vanish. Any remaining reducible gauge anomalies should be cancelled, e.g.~by a Higgs or Green-Schwarz (GS) mechanism. Terms in~$\CI^{\text{gauge}}_{d+2}$ that only involve background gauge fields encode 't Hooft anomalies for global symmetries, which need not cancel. Instead, they furnish robust observables that are often accessible even in strongly coupled theories. There can also be mixed anomalies involving dynamical and background gauge fields, which in general need not cancel either. 

The terms in~$\CI ^{\text{gravity}}_{d+2}$ are monomials in the Pontryagin classes~$p_k(T)$, which are~$4k$-forms in the curvature two-form~$R$. (The argument~$T$ refers to the tangent bundle.) If gravity is dynamical, then~$\CI ^{\text{gravity}}_{d+2}$, and any mixed anomalies involving gravity and dynamical gauge fields, must cancel. In this paper we will discuss quantum field theories; the metric only appears as a non-dynamical background field, and hence~$\CI ^{\text{gravity}}_{d+2}$ need not vanish.  Rather, the gravitational anomalies encoded by~$\CI ^{\text{gravity}}_{d+2}$ are analogous to 't Hooft anomalies for global symmetries, and the same is true for any mixed anomalies involving the gravity fields. 

In this paper we will discuss~$(1,0)$ SCFTs in six dimensions, which always possess an~$SU(2)_R$ symmetry. Therefore, the anomaly polynomial of such theories always contains terms of the form~\eqref{anomalyI}, which encode the~$SU(2)_R$ and gravitational anomalies of the theory. Throughout this paper we will collectively refer to them as the 't Hooft anomalies of the theory. (Some~$(1,0)$ theories also have flavor symmetries, which give rise to additional 't Hooft anomalies; they do not affect our discussion.) The~$SU(2)_R$ gauge field and its field strength will be denoted by~$A$ and~$F$, respectively. However, as in \eqref{anomalyI}, we use the notation~$c_2(R)$ rather than~$c_2(F)$ for the second Chern class of the~$R$-symmetry bundle. 

\subsection{Matching 't Hooft Anomalies on the Moduli Space}
\label{sec:moduli}

All known~$(1,0)$ SCFTs possess moduli spaces of vacua, where the conformal symmetry is spontaneously broken, even though Poincar\'e supersymmetry is preserved. The associated NG boson~$\varphi$ is known as the dilaton, and Goldstone's theorem implies that it becomes free in the deep IR. It must therefore reside in a free-field representation of~$(1,0)$ supersymmetry. In general, the moduli space may contain various branches, on which the dilaton can reside in different multiplets:

\begin{itemize}

\item[1.)] The tensor branch is parametrized by the expectation values of real scalars residing in~$(1,0)$ tensor multiplets. On such a branch the~$SU(2)_R$ symmetry is unbroken. The dilaton~$\varphi$ is the bottom component of one particular linear combination of the tensor multiplets whose scalars have acquired vevs. With the exception of a free hypermultiplet, all known~$(1,0)$ SCFTs possess a tensor branch. 

\item[2.)] The Higgs branch is parametrized by the expectation values of scalars~$q^i$ residing in~$(1,0)$ hypermultiplets, so that both the~$SU(2)_R$ symmetry and conformal symmetry are spontaneously broken. There are four NG bosons --   three for the~$SU(2)_R$ symmetry and one dilaton -- that reside in one particular linear combination of the hypermultiplets that have acquired a vev. The scalars in this NG hypermultiplet can be decomposed into a radial direction, corresponding to the dilaton, and an~$S^3$ of angular directions parametrized by the~$SU(2)_R$ NG bosons. 
\end{itemize}
\noindent There are also mixed branches, on which both tensor multiplets and hypermultiplets acquire vevs; we will not discuss them in detail. 

On both tensor and Higgs branches, there is typically a superficial mismatch between the 't Hooft anomalies of the massless fields in the IR and the SCFT in the UV, which is captured by the difference of the corresponding anomaly polynomials,
\begin{equation}
\Delta \CI_8 = \CI_8^{\text{UV}} - \CI_8^{\text{IR}}~.
\end{equation}
This mismatch is compensated by certain interactions involving the dynamical fields on the moduli space, as well as background fields, such as the metric or the~$R$-symmetry gauge field. (There is also a mismatch in the Weyl anomalies, which will be discussed in section~\ref{amatching}.) On the tensor branch, the 't Hooft anomalies are matched by GS-like interactions \cite{Green:1984bx,Sagnotti:1992qw} involving the two-form gauge fields residing in dynamical tensor multiplets, as well as background fields~\cite{Ohmori:2014kda,Intriligator:2014eaa}. On the Higgs branch, there are anomaly-matching interactions between the~$R$-symmetry NG bosons and background fields. We will now describe them in turn. 

As explained in~\cite{Ohmori:2014kda,Intriligator:2014eaa}, the 't Hooft anomalies on the tensor branch can be matched by a GS term in the effective Lagrangian, 
\begin{equation}
-i\Omega_{IJ} B^I\wedge X_4^J \subset~\SL~,\label{SBwedgeX}
\end{equation}
as long as the anomaly mismatch~$\Delta \CI_8$ is a sum of squares,
\begin{equation}
\Delta \CI _8 = \half \cdot {\Omega _{IJ} \over 2 \pi} \,  X_4^I\wedge X_4^J~.\label{dixsq}
\end{equation}
The factor of~$\half$ is due to the fact that the~$B^I$ are self-dual two-form gauge fields (the index~$I$ runs over all dynamical tensor multiplets); see appendix~\ref{app} for a more detailed discussion. For our purposes, the four-forms~$X_4^I$ will always be linear combinations of~$c_2(R)$ and~$p_1(T)$. The matrix~$\Omega _{IJ}$ in~\eqref{SBwedgeX} and~\eqref{dixsq} is symmetric and positive definite. It determines the Dirac pairing between self-dual string sources that couple to the two-form gauge fields~$B^I$ residing in the tensor multiplets,\footnote{~Since the~$X_4^I$ act as sources for the tensors~$B^I$, they are constrained by Dirac quantization~\cite{Ohmori:2014kda,Intriligator:2014eaa}. These quantization conditions take a particularly simple form if one chooses a non-canonical normalization for the tensor multiplets (see for instance equation~(1.8) in~\cite{Intriligator:2014eaa}), which differs from the one used here.} as well as the kinetic terms of their superpartners. (Since the~$B^I$ have self-dual field strengths, they do not possess meaningful kinetic terms.) It will be convenient to work in a basis in which the tensor-multiplet scalars have canonically normalized kinetic terms, so that~$\Omega_{IJ} = \delta_{IJ}$.  (This differs from the normalizations in~\cite{Ohmori:2014kda,Intriligator:2014eaa}; in particular, our equation~\eqref{dixsq} contains an additional factor of~$2 \pi$.) On a rank one tensor branch described by a single tensor multiplet we have~$\Omega = 1$ and
\begin{equation}\label{xydef}
X_4 = 16\pi^2 \left(x \, c_2(R) + y \, p_1(T) \right)~,
\end{equation}
where~$x, y$ are real coefficients. (The prefactor is chosen for later convenience and will be explained in section~\ref{sugra}.) Substituting into~\eqref{dixsq} and comparing with the general form of the anomaly polynomial in~\eqref{anomalyI} then leads to~\eqref{rk1square}.  Note that the irreducible gravitational anomaly~$p_2(T)$ cannot be matched by GS mechanism, and hence it must take the same value in the UV and IR theories. 

On the Higgs branch, a GS mechanism is not available and all anomalies must be absorbed using the~$SU(2)_R$ NG bosons. Therefore the anomaly mismatch must be of the form
\begin{equation}
\Delta \CI _8=c_2(R) \wedge X_4~,
\end{equation}
for some four-form~$X_4$. This involves neither the irreducible gravitational anomaly~$p_2(T)$, nor the reducible one~$p_1^2(T)$. Therefore the UV anomaly coefficients~$\gamma$ and~$\delta$ in~\eqref{anomalyI} can be expressed in terms of the quaternionic dimension~$d_{\text{Higgs}}$ of the Higgs branch using the anomaly coefficients of a free hypermultiplet (see Table~\ref{known}),
\begin{equation}\label{dimhiggs}
\gamma = -{7 \over 240} \, d_{\text{Higgs}}~, \qquad \delta = {1 \over 60} \, d_{\text{Higgs}}~.
\end{equation}
Conversely, a~$(1,0)$ SCFT can only admit a pure Higgs branch if its gravitational anomaly coefficients can be expressed as~\eqref{dimhiggs} for some positive integer~$d_{\text{Higgs}}$.

\subsection{The Dilaton Effective Lagrangian and the~$a$-Anomaly}
\label{amatching}

Since conformal symmetry is spontaneously broken on the moduli space of~$(1,0)$ SCFTs, the low-energy theory always contains a weakly interacting massless scalar -- the dilaton -- which is the NG boson of conformal symmetry breaking. We will now review the structure of the dilaton effective Lagrangian~$\SL_{\text{dilaton}}$ and its relation to the~$a$-type Weyl anomaly. The constraints of supersymmetry will be explored in subsequent sections. 

Following the work of~\cite{Komargodski:2011vj,Komargodski:2011xv} in four dimensions, the constraints of non-linearly realized conformal symmetry on the low-energy effective Lagrangian of the dilaton~$\varphi$ were analyzed in~\cite{Maxfield:2012aw, Elvang:2012st,Elvang:2012yc}. This analysis is facilitated by coupling the dilaton to a background metric~$g_{\mu\nu}$.  Under a local Weyl rescaling, the dilaton and the metric transform as follows,
\begin{equation}\label{Weylresc}
\varphi \rightarrow  e^{-2 \sigma} \varphi~, \qquad g_{\mu\nu} \rightarrow e^{2 \sigma} g_{\mu\nu}~.
\end{equation}
It is convenient to define the Weyl-invariant combination~$\hat g_{\mu\nu} = {\varphi \over \langle \varphi\rangle} g_{\mu\nu}$, where~$\langle \varphi\rangle$ is the dilaton vev. 

All local curvature invariants of~$\hat g_{\mu\nu}$ lead to acceptable terms in~$\SL_{\text{dilaton}}$, once we choose a flat background metric~$g_{\mu\nu} = \delta_{\mu\nu}$. For instance, the Einstein-Hilbert term for~$\hat g_{\mu\nu}$ induces a dilaton kinetic term~$\half \left(\d\varphi\right)^2 \subset \SL_{\text{dilaton}}$. The four-derivative terms in the dilaton Lagrangian arise from the contraction of two Ricci tensors,
\begin{equation}
\langle \varphi  \rangle \sqrt{\hat g} \, \hat R_{\mu\nu} \hat R^{\mu\nu} \quad  \longrightarrow \quad - \half {\left(\d\varphi\right)^4 \over \varphi^3}~.\label{dilaton4dsec2}
\end{equation}
Other curvature-squared terms give rise to terms that do not affect the flat-space dilaton Lagrangian, and hence we will not consider them.

At the six-derivative order, conformal symmetry requires a very particular dilaton interaction term of the following schematic form (see~\cite{Elvang:2012st} for a detailed discussion),
\begin{equation}
\Delta a \, \sqrt{-g} \, \log \varphi \,  E_6 \quad \longrightarrow \quad \Delta a \, {\left(\d \varphi\right)^6 \over \varphi^6}~.
\label{amatch}
\end{equation}
Here~$E_6$ is the Euler density and~$\Delta a = a_{\text{UV}} - a_{\text{IR}}$ is the mismatch between the~$a$-type Weyl anomalies of the UV and IR theories. The Wess-Zumino-like term in~\eqref{amatch} is needed to absorb this mismatch, and it leads to a non-trivial six-derivative term for the dilaton even if the background metric is flat. Below, we will heavily rely on the fact that the~$a$-anomaly appears in the flat-space effective action on the moduli space of~$(1,0)$ SCFTs.

We can summarize the preceding discussion by writing the first few terms in the dilaton effective Lagrangian,
\begin{equation}
\SL_{\text{dilaton}} = \frac{1}{2}\left(\partial \varphi\right)^{2}-b\, \frac{\left(\partial \varphi\right)^{4}}{\varphi^{3}} + \Delta a\,\frac{\left(\partial \varphi\right)^{6}}{\varphi^{6}}+\CO\left(\d^8\right)~, \label{ldilaton}
\end{equation}
where the expression for the six-derivative term is schematic, as in~\eqref{amatch}. The constant~$b$ is a dimensionless coupling, whose definition is tied to the canonical choice of kinetic terms in~\eqref{ldilaton}. Following~\cite{Elvang:2012st}, it is useful to note that~$b$ determines the~$\CO(p^4)$ on-shell scattering amplitude of four dilatons (here~$p$ denotes the overall momentum scale), which does not suffer from field-redefinition ambiguities. A dispersion relation for this amplitude shows that~$b > 0$ unless the dilaton is a free field, in which case~$b$ vanishes~\cite{Adams:2006sv} (see also~\cite{Komargodski:2011vj,Komargodski:2011xv,Elvang:2012st}). Similarly, $\Delta a$ appears at~$\CO(p^6)$ in dilaton scattering amplitudes. 

As long as the theory in the deep IR is a conventional CFT, we can treat~$\varphi$ as a decoupled field, which only interacts with itself, up to and including six-derivative order. (See~\cite{Komargodski:2011xv,Luty:2012ww} for a discussion of the corresponding statement in four dimensions.) At higher orders in the derivative expansion, we must also take into account possible couplings of~$\varphi$ to other massless degrees of freedom in the IR, which can give rise to non-analytic terms in dilaton scattering amplitudes. As was stated in the introduction, many~$(1,0)$ SCFTs admit RG flows that do not terminate in conventional CFTs. These are discussed in section \ref{sec:vectors}.

\section{The~$a$-Theorem for Tensor Branch Flows}
\label{sec:athm}

In this section, we analyze the low-energy effective action for the dilaton~$\varphi$ on the tensor branch of a six-dimensional~$(1,0)$ SCFT~$\CT_{\text{UV}}$. We show that supersymmetry implies that the coefficients in~\eqref{ldilaton} satisfy a universal relation of the form~$\Delta a \sim b^2$, with a positive, model-independent proportionality constant. Moreover, as a consequence of unitarity, the coefficient $b$ in \eqref{ldilaton} satisfies~\cite{Elvang:2012st}
\begin{equation}
b\geq 0~,
\end{equation}
with equality in the above if and only if the dilaton is free and the associated RG flow is trivial.  So ~$\Delta a \sim b^2$ implies that~$\Delta a > 0$ unless the flow is trivial, in which case~$\Delta a = 0$, thus proving~$a$-theorem for RG flows of~$(1,0)$ SCFTs onto their tensor branch. 

Throughout our discussion here, we assume that the massless degrees of freedom that remain in the deep IR for the theory with non-zero expectation value on the tensor branch constitute a genuine SCFT~$\CT_{\text{IR}}$. Since the dilaton~$\varphi$ is the NG boson of spontaneous conformal symmetry breaking, it follows from Goldstone's theorem that~$\CT_{\text{IR}}$ consists of a (possibly interacting) SCFT~$\CT_0$ and a free decoupled tensor multiplet~$\CT_\varphi$ containing the dilaton~$\varphi$,
\begin{equation}
\CT_{\text{IR}} = \CT_0 + \CT_{\varphi}~.
\end{equation}

The assumption that~$\CT_0$ is an SCFT excludes tensor branches with massless gauge fields. If the IR theory is not a CFT, both the statement and the proof of an~$a$-theorem require additional clarification. For this reason we defer a discussion of tensor branches with gauge fields until section \ref{sec:vectors}. Prototypical examples of SCFTs without gauge fields on their tensor branch are the~$(2,0)$ theories, as well as the~$(1,0)$ theories~$\CE_N$ describing~$N$ small~$E_8$ instantons, which will be discussed in section \ref{sec:E8}.

We now turn to the implications of~$(1,0)$ supersymmetry for the dilaton effective action~\eqref{ldilaton} on the tensor branch. Here we closely follow the recent discussion of tensor-branch effective actions with~$(2,0)$ supersymmetry in~\cite{Cordova:2015vwa}. As was explained in section~\ref{sec:anomaly}, the dilaton~$\varphi$ resides in a~$(1,0)$ tensor multiplet, together with a symplectic Weyl Fermion~$\psi_\alpha^i$ and a self-dual three-form field strength~$H$, which can be written as a symmetric bispinor~$H_{\alpha\beta} = H_{(\alpha\beta)}$. (Here~$\alpha, \beta = 1, \ldots, 4$ are chiral spinor indices and~$i = 1,2$ is an~$SU(2)_R$ doublet index.) At the two-derivative level, they all satisfy free equations of motion,
\begin{equation}\label{freeeom}
\square \varphi = 0~, \qquad \d^{\alpha\beta} \psi_\beta^i = 0~, \qquad \d^{\alpha\beta} H_{\beta\gamma} = 0~.
\end{equation}
Here~$\d^{\alpha\beta} = \d^{[\alpha\beta]}$ is a spacetime derivative in bispinor notation. The fact that~$H$ is self dual implies that the standard quadratic Lagrangian~$H \wedge * H$ vanishes, so that the free theory needs to be defined with some care (see for instance~\cite{Witten:2007ct,Witten:2009at,moorefklect} and references therein), but this subtlety will not affect our discussion.

Since the tensor-branch effective action is supersymmetric, the higher-derivative terms in the pure dilaton Lagrangian~\eqref{ldilaton} for~$\varphi$ must be completed by terms involving its superpartners~$\psi_\alpha^i$ and~$H_{\alpha\beta}$.\footnote{~We follow the standard rules for counting derivatives in supersymmetric moduli-space effective actions: spacetime derivatives and the three-form field strength~$H$ have weight~$1$, supercharges and Fermions have weight~$\half$, and the scalar~$\varphi$ has weight~$0$.} In order to determine whether supersymmetry leads to additional constraints on these terms, we follow the general approach to moduli-space effective actions advocated in~\cite{Cordova:2015vwa,Cordova:2016xhm}. We first expand the dilaton in fluctuations~$\delta \varphi$ around a fixed vev,
\begin{equation}\label{vevfluc}
\varphi = \langle \varphi \rangle + \delta \varphi~,
\end{equation}
and view the resulting Lagrangian as a deformation of a free tensor multiplet by higher-derivative local operators constructed out the fields in this tensor multiplet and their derivatives. If some term in this expansion leads to local operators that cannot be embedded in an independent supersymmetric deformation, then that term is constrained by supersymmetry, i.e.~it satisfies a non-renormalization theorem.  

To implement this procedure, we now determine the independent supersymmetric deformations of a single free~$(1,0)$ tensor multiplet. Since this multiplet constitutes a (free) SCFT, its supersymmetric deformations can be classified using superconformal representation theory, see \cite{Cordova:2016xhm,Cordova:2016emh} and references therein for further details. (Below we will mention another approach, based on scattering superamplitudes.) Unlike the~$(2,0)$ case discussed in~\cite{Cordova:2015vwa}, which admits both~$F$- and~$D$-term deformations, a free~$(1,0)$ tensor multiplet can only be deformed by full~$D$-terms, i.e. descendants formed with all supercharges
\begin{equation}
\SL_{D}=Q^{8}\left(\mathcal{O}\right)~,\label{dterm}
\end{equation}
where~$\CO$ is constructed out of fields in the tensor multiplet and their derivatives. This is similar to the situation in four-dimensional~$\CN=2$ theories, where all higher-derivative operators on the Coulomb branch are full~$D$-terms~\cite{Dine:1997nq}.  In order for the deformation~\eqref{dterm} to be non-trivial, $\CO$ must be the superconformal primary (i.e.~the bottom component) of a long multiplet, which does not satisfy any shortening conditions. Both~$\CO$ and~$\SL_D$ must be Lorentz scalars, and they transform in the same representation of the~$SU(2)_R$ symmetry; the eight supercharges in~\eqref{dterm} are contracted to an~$SU(2)_R$ singlet.  The $SU(2)_R$ symmetry is unbroken on the tensor branch, so the operator~$\CO$ must be an~$SU(2)_R$ singlet.  

The leading interaction in the dilaton effective Lagrangian~\eqref{ldilaton} is the four derivative term, proportional to~$b$. Expanding it around a fixed dilaton vev as in \eqref{vevfluc} gives rise to an infinite series of four-derivative terms involving~$n + 4$ dilatons, for all ~$n \in \Z_{\geq 0}$,
\begin{equation}
b \, \frac{\left(\partial \varphi\right)^{4}}{\varphi^{3}} \quad \longrightarrow \quad b \left(\frac{1}{\langle \varphi\rangle ^{3}} (\partial \delta \varphi)^{4} -3 \,\frac{\delta \varphi}{\langle \varphi \rangle^{4}}(\partial \delta \varphi)^{4} + \cdots +\CO\left((\delta \varphi)^n (\d \delta \varphi)^4\right) + \cdots \right)~.\label{fourderivative}
\end{equation}
All terms in this expansion can be interpreted as arising from~$D$-terms \eqref{dterm} of the form $Q^8\left((\delta \varphi)^{n+4}\right)$.  Therefore, supersymmetry does not constrain the coefficient~$b$. 

We now note that the six-derivative couplings in~\eqref{ldilaton}, which are proportional to~$\Delta a$, cannot arise from a~$D$-term~\eqref{dterm}. In order to see this, it suffices to list all candidate 
Lorentz and~$SU(2)_R$ singlet primaries~$\CO$, which should contain two derivatives for~\eqref{dterm} to be a six-derivative term.  However, no such~$\CO$ exists, as can be verified by enumerating all local Lorentz scalar operators containing two derivatives: 
\begin{itemize}

\item $(\delta\varphi)^{n}(\partial \delta\varphi)^{2}  $ is not a conformal primary (i.e.~it is a total derivative), since~$\square \varphi = 0$.

\item $\psi_{\alpha}^{i}\psi_{\beta}^{j}\partial^{\alpha\beta}(\delta\varphi^{n})$ transforms in the~$\bf 3$ of ~$SU(2)_R$, since~$\d^{\alpha\beta} = \d^{[\alpha\beta]}$ is antisymmetric.

\item $(\delta \varphi)^{n}\varepsilon^{\alpha \beta \gamma \delta}\psi_{\alpha}^{i}\psi_{\beta}^{j}\psi_{\gamma}^{k}\psi_{\delta}^{\ell}$ transforms in the~$\bf 5$ of~$SU(2)_R$, since the totally antisymmetric~$\ep$-symbol is needed to contract the spinor indices to a Lorentz singlet. 
\item $H_{\alpha \beta}\partial^{\alpha \beta}\left((\delta\varphi)^{n}\right)=(\delta \varphi)^{n} \ep^{\alpha\beta\gamma\delta} \psi_{\alpha}^{i}\psi_{\beta}^{j}H_{\gamma\delta} =  (\delta \varphi)^{n}\ep^{\alpha\beta\gamma\delta} H_{\alpha \beta}H_{\gamma\delta} = 0$ since~$H_{\alpha\beta} = H_{(\alpha\beta)}$ is symmetric and~$\d^{\alpha\beta} = \d^{[\alpha\beta]}$ is antisymmetric.
\end{itemize}

The absence of independent six-derivative couplings implies that the six-derivative terms in~\eqref{ldilaton}, and their superpartners, can only be induced by supersymmetrically completing the four-derivative terms~\eqref{fourderivative} proportional to~$b$. Thus all six-derivative terms in the dilaton effective Lagrangian must be proportional to~$b^2$. See~\cite{Cordova:2015vwa} for various ways of understanding this quadratic relation. Therefore the coefficient $\Delta a$ of the six-derivative term in~\eqref{ldilaton} must be proportional to~$b^2$, with a model-independent proportionality constant that is completely fixed by supersymmetry. This coefficient can be  determined directly, or via examining any suitable example. For instance, it was shown in the~\cite{Cordova:2015vwa} that
\begin{equation}\label{absq}
\Delta a = {98304 \pi^3 \over 7} b^2~,
\end{equation}
for all~$(2,0)$ SCFTs. Since these are also particular examples of~$(1,0)$ SCFTs, the relation~\eqref{absq} continues to hold on the tensor branch of all~$(1,0)$ theories, due to the non-renormalization theorem derived above.\footnote{~It is interesting to contemplate the extent to which a relation like~\eqref{absq}, which was derived using spontaneously broken conformal symmetry, as well as supersymmetry, continues to apply if we relax some of these assumptions. See for instance the discussion around equation~(2.7) in~\cite{Maldacena:1997re}. We thank J.~Maldacena for suggesting this possibility, and for related discussions.}

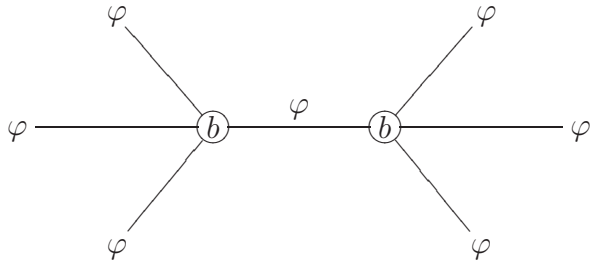
\begin{figure}[h]
\[
\xymatrix  @R=1pc {
&*{\varphi}\ar@{-}[ddr]&&&&   *{~\varphi}\ar@{-}[ddl]&\\
\\
{\varphi}\ar@{-}[rr]&&*+[o][F]{ b } \ar@{-}[rr]^{\mathlarger{\varphi}}&&   *+[o][F]{ b }\ar@{-}[l]&&{\varphi}\ar@{-}[ll]\\
 \\
&{\varphi}\ar@{-}[uur]&&&&   {\varphi}\ar@{-}[uul]&}  \]

 \caption{Factorization of a six-point dilaton amplitude proportional to~$\Delta a$ through a pair of four-point amplitudes proportional to~$b$. This explains the quadratic relation~$\Delta a \sim b^2$.} 
 \label{Factor}
\end{figure}

As in~\cite{Cordova:2015vwa}, the fact that~$\Delta a$ is proportional to~$b^2$ can also be understood by examining tree-level scattering amplitudes of the fields in the dilaton multiplet. The implications of~$(1,0)$ supersymmetry on local tensor multiplet supervertices were analyzed in~\cite{Chen:2015hpa}, where it was shown that these constraints are incompatible with the existence of an independent six-point, six-derivative supervertex. Any six-point, six-derivative treel-level amplitude of fields in the dilaton multiplet must therefore factorize through a product of four-point, four-derivative amplitudes (see Figure~\ref{Factor}), which again leads to the quadratic relation~$\Delta a \sim b^2$. Similar techinques have recently been used to argue for supersymmetry relations between various higher-derivative couplings in diverse dimensions~\cite{Wang:2015jna,Lin:2015ixa,Wang:2015aua}, including those previously obtained using the methods of~\cite{Paban:1998qy,Paban:1998ea}.

The universal quadratic relation in~\eqref{absq} immediately implies the~$a$-theorem for tensor-branch flows. As was reviewed above, $b > 0$ unless the dilaton is a free field, in which case~$b = 0$. Therefore~$\Delta a>0$ for all non-trivial RG flows of~$(1,0)$ SCFTs onto their tensor branch. This argument for the~$a$-theorem is identical to that given in~\cite{Cordova:2015vwa} for tensor-branch flows of~$(2,0)$ theories (see also~\cite{Maxfield:2012aw}), since the crucial relation~\eqref{absq} is the same in both cases.

\section{Relating the~$a$-Anomaly to 't Hooft Anomalies}
\label{sec:relation}

In this section we derive formula~\eqref{ais}, which relates the~$a$-anomaly to the coefficients~$\alpha, \beta, \gamma, \delta$ that appear in the anomaly polynomial~\eqref{anomalyI}. As in the previous section, we consider RG flows onto the tensor branch, where the dilaton is in a tensor multiplet.  We show that supersymmetry relations among the anomaly-matching interactions imply that the changes in the anomaly coefficients along such flows must satisfy
\begin{equation}\label{sectiongoal}
\Delta a = {16 \over 7} \left(\Delta \alpha - \Delta \beta + \Delta \gamma\right)~.
\end{equation}
In order to establish this relation, we couple the theory to background conformal supergravity fields. This reveals a universal linear relation, required by superconformal symmetry, between the coefficients of the GS couplings in~\eqref{SBwedgeX} and the coefficient~$b$ of the four-derivative dilaton interaction in~\eqref{ldilaton}. The former are quadratically related to~$\Delta \alpha, \Delta \beta, \Delta \gamma$ by the GS mechanism, as in~\eqref{rk1square}, while~$b$ is quadratically related to~$\Delta a$ by supersymmetry, as in~\eqref{absq}, and together these lead to~\eqref{sectiongoal}. This only leaves the coefficient of~$\delta$ in~\eqref{ais} undetermined, which can be fixed by examining the anomalies of a free hypermultiplet. 

\subsection{Rank One Tensor Branches}
\label{sugra}

We first consider a rank one tensor branch, which is described by a single tensor multiplet.  We show that the coefficient~$b$ in~\eqref{ldilaton} is given by a particular linear combination of the coefficients~$x, y$ in~\eqref{xydef}, which determine the GS term~\eqref{SBwedgeX} that is needed to match the~$SU(2)_R$ and gravitational anomalies on the tensor branch. 

The GS term~\eqref{SBwedgeX} involves the background metric and background~$SU(2)_R$ gauge field.  To supersymmetrize these interactions, in the spirit of~\cite{Festuccia:2011ws}, these fields should be embedded in a rigid, background supergravity multiplet. Since the dilaton effective action is superconformal, the appropriate choice is the~$(1,0)$ superconformal gravity multiplet constructed in~\cite{Bergshoeff:1985mz} and further explored in~\cite{Bergshoeff:1986vy,Nishino:1986dc,Bergshoeff:1986wc,Bergshoeff:1987rb}   (see~\cite{Coomans:2011ih} for a recent discussion). We will mostly rely on~\cite{Bergshoeff:1986wc}, which uses the conventions of~\cite{Bergshoeff:1985mz}.  The independent fields that describe the tensor multiplet containing the dilaton coupled to conformal supergravity are given by\footnote{~This field content is described in appendix~C of~\cite{Bergshoeff:1985mz} and section~3 of~\cite{Bergshoeff:1986wc}. We follow their conventions, up to the following changes in notation: $\varphi_{\text{us}} = \sigma_{\text{them}}$ and~$\big(A_\mu^{ij}\big)_{\text{us}} = -{ i \over 2} \big(V_\mu^{ij}\big)_{\text{them}}$.}
\begin{equation}\label{gravfields}
\left(\, \varphi, \; \psi^i_\alpha~,\; B_{\mu\nu}~, \; g_{\mu\nu}~, \; \psi_{\mu}^{\alpha i}~,\; A_{\mu}^{ij}\, \right)~. 
\end{equation}
The dilaton~$\varphi$ and the fermion~$\psi_\alpha^i$ reside in the $\CN =(1,0)$ tensor multiplet. The field strength of the two-form~$B_{\mu\nu}$ has both a self-dual and an anti self-dual part. Roughly speaking, the self-dual part can be identified with the self-dual three-form field strength~$H$ of the $(1,0)$ tensor multiplet, to which it reduces when the conformal supergravity fields are set to zero. The anti-self-dual part of ~$B_{\mu\nu}$ resides in the background gravity multiplet, together with the metric ~$g_{\mu\nu}$, gravitino~$\psi_\mu^{\alpha i}$, and the~$SU(2)_R$ gauge field $A_\mu^{ij} = A_\mu^{(ij)}$.  Here ~${A^i}_j = {A_\mu^i}_j dx^\mu$ is a traceless Hermitian matrix in the fundamental representation of~$SU(2)_R$, with field strength~${F^i}_j = \half {F^i}_{j \mu\nu} dx^\mu \wedge dx^\nu$ given by 
\begin{equation}
F = dA - i A \wedge A~.
\end{equation}
The~$SU(2)_R$ indices are raised and lowered with a two-index~$\ep$-symbol according to the conventions of~\cite{Bergshoeff:1985mz}. Since the second Chern class $c_2(R)$ is normalized so that it integrates to~$1$ on a minimal~$SU(2)_R$ instanton in flat space (more precisely on~$S^4$), we have
\begin{equation}
c_2(R) = {1 \over 8 \pi^2} \tr\left(F \wedge F\right)~, 
\end{equation}
where~$\tr$ denotes the trace in the fundamental representation of~$SU(2)_R$, i.e.~over the matrix indices of~${F^i}_j$. 

Whenever the dilaton~$\varphi$ has a non-zero vev, as is the case on the tensor branch, we can set~$\varphi = \langle \varphi\rangle$ to a constant by a local Weyl rescaling. Equivalently, we can follow the discussion in section \ref{amatching} and define a Weyl-invariant metric~$\hat g_{\mu\nu} = {\varphi \over \langle \varphi\rangle} g_{\mu\nu}$, which is equal to~$g_{\mu\nu}$ in the gauge~$\varphi = \langle \varphi\rangle$. We can similarly remove the fermion~$\psi_\alpha^i$ by gauge fixing the local special conformal transformations. The transformation rules for the remaining fields~$B_{\mu\nu}, A_\mu^{ij}, g_{\mu\nu}, \psi_\mu^{\alpha i}$ are modified in order to preserve these gauge choices. We can then write Lagrangians that are invariant under local supersymmetry transformations, diffeomorphisms, and~$SU(2)_R$ gauge transformations. The dependence on the dilaton is easily restored by performing a local Weyl rescaling~\eqref{Weylresc} with parameter~$\sigma \sim \log \varphi$.

Fortuitously, the needed supergravity completion of the GS term~\eqref{SBwedgeX} was already worked out long ago, in the context of six-dimensional~$R^2$ supergravity~\cite{Bergshoeff:1986vy,Bergshoeff:1986wc,Bergshoeff:1987rb}. There are two independent terms, corresponding to the two coefficients~$x, y$ appearing in $X_4$ in ~\eqref{xydef}. They can be found in equations~(B.1) and~(C.1) of~\cite{Bergshoeff:1986wc}, in the gauge where~$\varphi = \langle \varphi \rangle $ is a constant. Here we will only display those terms that will be important for us: the pure curvature-squared terms, and the GS terms, which in our notation are
\begin{subequations}
\begin{align}
& \SL_{R^2} = \langle \varphi \rangle \sqrt g \, \left( \Big(y - {x \over 4}\Big) \,R^{\mu\nu\rho\lambda} R_{\mu\nu\rho\lambda} + {3 \over 2} x \,{R_{[\mu\nu}}^{\mu\nu} {R_{\rho\sigma]}}^{\rho\sigma}\right)~,\label{rsq}\\
& \SL_{\text{GS}} = 16 i  \pi^2 \, B \wedge \left(x \,  c_2(R) + y \, p_1(T)\right)~.\label{gslag}
\end{align}
\end{subequations}
Here~$R_{\mu\nu\rho\lambda}$ is the Riemann curvature tensor (we follow the curvature conventions explained in section~2 of~\cite{Bergshoeff:1986wc}) and we have used the fact that the first Pontryagin class is
\begin{equation}
p_1(T) = {1 \over 8 \pi^2} \tr \left(R \wedge R\right)~,
\end{equation}
where~$\tr$ is a trace over~$SO(6)$ tangent frame indices. 

As was explained around~\eqref{dilaton4dsec2} (see also section 3.2 of~\cite{Elvang:2012st}), the four-derivative dilaton coupling with coefficient~$b$ only arises from the contraction of two Ricci tensors; we can drop contributions involving the Weyl tensor~${W_{\mu\nu}}^{\alpha\beta}$ or the Ricci scalar~$R$ from the Riemann tensor
\begin{equation}
{R_{\mu\nu}}^{\alpha\beta} = {W_{\mu\nu}}^{\alpha\beta} + \delta^{[\alpha}_{[\mu} R^{\beta]}_{\nu]} -{1 \over 10} \delta^\alpha_{[\mu} \delta^\beta_{\nu]} \, R~.
\end{equation}
Substituting this into~\eqref{rsq} and taking the flat space limit then leads to 
\begin{equation}\label{dilaton4d}
\SL_{R^2} \quad \longrightarrow \quad - \half \left(y -x \right) {\left(\d \varphi\right)^4 \over \varphi^3}~.
\end{equation}
This establishes our desired relation between the dilaton and tensor interactions: 
\begin{equation}\label{bxyeq}
b = \half \left(y - x\right)\geq0~.
\end{equation}
Unitarity requires that the overall sign of~$x$ and~$y$, which is undetermined by the GS mechanism, should be chosen so that~$b \geq 0$~\cite{Elvang:2012st}.

The coefficients~$x$ and~$y$ in~\eqref{gslag} give the GS contribution~\eqref{dixsq} and~\eqref{xydef},
\begin{equation}
\Delta \CI_8 = {1 \over 4 \pi}X^2_4 = 64 \pi^3 \left(x^2 c_2^2(R) + 2 x y \, c_2(R) p_1(T) + y^2  p_1^2(T)\right)~.
\end{equation}
So~$x$ and~$y$ are related to the changes in the anomaly coefficients~\eqref{anomalyI} as follows, 
\begin{equation}\label{abcxy}
\Delta \alpha = 1536 \pi^3 x^2~, \qquad \Delta \beta = 3072 \pi^3 x y~, \qquad \Delta \gamma = 1536 \pi^3 y^2~.
\end{equation}
Substituting~\eqref{bxyeq} into~\eqref{absq} we find that
\begin{equation}
\Delta a = {24 576 \pi^3 \over 7} \left(x-y\right)^2~. \label{deltaaxy}
\end{equation}
Using~\eqref{abcxy}, this finally leads to\begin{equation}\label{deltafinal}
\Delta a = {16 \over 7} \left(\Delta \alpha - \Delta \beta + \Delta \gamma\right)~.
\end{equation}

\subsection{Tensor Branches of Higher Rank}
\label{sec:highrk}

We may reach a general point on a higher dimensional tensor branch by a sequence of rank one flows, each as in the previous subsection.  This implies that
\begin{equation}
\Delta \alpha = 1536 \pi^3 \vec x \cdot \vec x ~, \qquad \Delta \beta = 3072 \pi^3 \vec x \cdot \vec y~, \qquad \Delta \gamma = 1536 \pi^3 \vec y\cdot \vec y~
\end{equation}
where $\vec x\cdot \vec y\equiv \sum_{I,J}\Omega _{IJ}x^I y^J$, summed over all tensor multiplets,  and 
\begin{equation}
\Delta a = {24 576 \pi^3 \over 7} \left(\vec x-\vec y\right)^2~. 
\end{equation}
This shows that the same relation~\eqref{deltafinal} still holds. Since~$\Omega_{IJ}$ is positive definite, the conclusion that~$\Delta a\geq 0$ also remains valid.

\subsection{A Universal Formula for the~$a$-Anomaly}
\label{sec:finalform}
The relation~\eqref{deltafinal} for the changes of the anomalies on the tensor branch  implies that a universal linear relation between the~$a$-anomaly and the 't Hooft anomaly coefficients~$\alpha, \beta, \gamma, \delta$ must take the form
\begin{equation}\label{aansatz}
a = {16 \over 7} \left(\alpha- \beta + \gamma\right) + K \delta~.
\end{equation}
The~$K\delta$ term drops out in $\Delta a$ because $\Delta \delta =0$ everywhere on the moduli space.  The constant $K$, which we expect is also fixed by supersymmetry, can be determined by  evaluating both sides of~\eqref{aansatz} for any known example SCFT with $\delta \neq 0$.  For example, for a free~$(1,0)$ hypermultiplet (see Table~\ref{known})~$a = {11 \over 210}$, $\alpha = \beta = 0$, $\gamma = {7 \over 240}$, and~$\delta = -{1 \over 60}$. Substituting into~\eqref{aansatz} gives~$K = {6 \over 7}$, which leads to our formula~\eqref{ais} for the~$a$-anomaly:
\begin{equation}
a = {16 \over 7} \left(\alpha- \beta + \gamma\right) + {6 \over 7} \, \delta~.  \label{ais2}
\end{equation}
It is a non-trivial check that this formula is also consistent with the anomaly coefficients of a free~$(1,0)$ tensor multiplet and all~$(2,0)$ SCFTs, as summarized in Table~\ref{known}.

\section{Example: The Theory of~$N$ Small~$E_8$ Instantons}
\label{sec:E8}

\begin{figure}[h]
\[
\xymatrix  @R=1pc {
&&&&*=0{\phantom{\bullet}}\ar@{--}[ddddddddd] &&&&&  &&  *=0{\phantom{\bullet}}\ar@{--}[ddddddddd]&&\\
&&*=0{\phantom{\bullet}}\ar@{-}[ddddd]   &&&&   &&*=0{\phantom{\bullet}}\ar@{-}[ddddd]&&    &&&  &&*=0{\phantom{\bullet}}\ar@{-}[ddddd]  \\
\\
*=0{\phantom{\bullet}}\ar@{-}[uurr]\ar@{-}[ddddd]&& &&&&  *=0{\phantom{\bullet}}\ar@{-}[uurr]\ar@{-}[ddddd]&&   &&   &&& *=0{\phantom{\bullet}}\ar@{-}[uurr]\ar@{-}[ddddd]&& \\
&&&&&&&&&&&&&& *=0{\bullet } & \\
& *=0{\bullet } & &&&& & *=0{\bullet} & & *=0{\bullet}&  &&&  & *=0{\bullet } & \\
&*=0{ N}&*=0{\phantom{\bullet}}&&&&  &*=0{ N-1}&*=0{\phantom{\bullet}} && &&&  &*=0{\bullet }&*=0{\phantom{\bullet}} \\
\\
*=0{\phantom{\bullet}}\ar@{-}[uurr]&&  &&&&  *=0{\phantom{\bullet}}\ar@{-}[uurr]&& && &&& *=0{\phantom{\bullet}}\ar@{-}[uurr]&&  \\
& *=0{(a)} &  &&*=0{\phantom{\bullet}}&   &&& *=0{(b)} &&& *=0{\phantom{\bullet}} && &*=0{(c)}}  \]

 \caption{M-Theory description of the~$(1,0)$ SCFT~$\CE_N$ of~$N$ small~$E_8$ instantons. In (a) there are $N$ M5-branes (represented by dots) embedded in the Ho\v{r}ava-Witten wall.  In (b) a flow onto the tensor branch is initiated by pulling a single M5-brane off the wall.  In (c) a Higgs branch flow is initiated by dissolving the branes inside the wall.}
  \label{figE8}
 \end{figure}
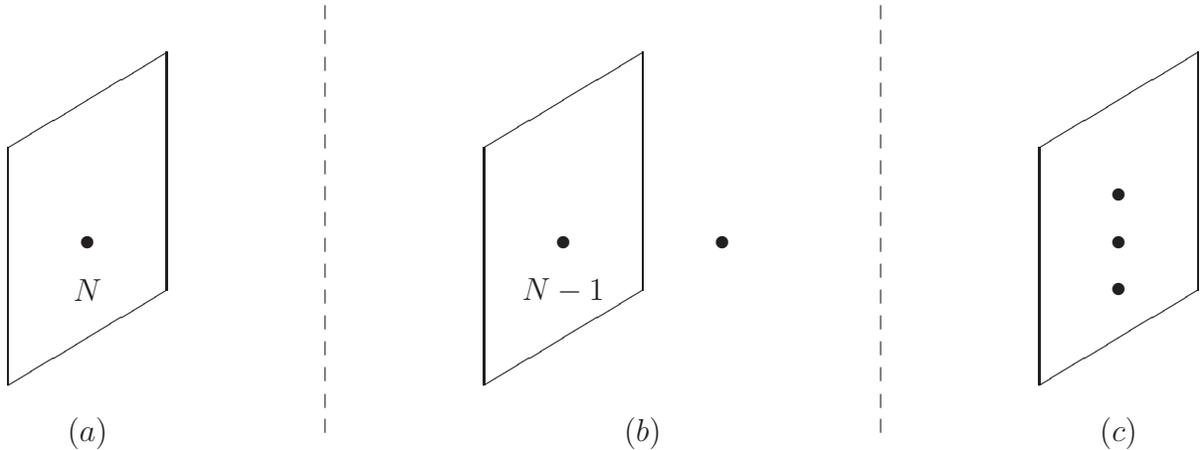

We can now use our formula~\eqref{ais2} to compute the~$a$-anomaly for~$(1,0)$ SCFTs whose 't Hooft anomalies are known, and to study RG flows between such theories. In this section, we consider the SCFT~$\CE_N$ on the worldvolume of~$N$ small, coincident $E_{8}$ instantons in heterotic string theory~\cite{Ganor:1996mu,Seiberg:1996vs}. From the M-theory viewpoint, the theory~$\CE_N$ arises when~$N$ coincident M5-branes are embedded in the Ho\v{r}ava-Witten wall \cite{Horava:1996ma}, as illustrated in Figure~\ref{figE8}(a). It is convenient to include the center of mass mode of the M5-branes, which is described by a free hypermultiplet, in the definition of~$\CE_N$.  The anomaly polynomial of~$\CE_N$ was determined in~\cite{Ohmori:2014pca,Ohmori:2014kda},
\begin{equation}
\alpha= N(4N^{2}+6N+3)~, \qquad \beta= -\frac{N}{2}(6N+5)~,\qquad \gamma= \frac{7N}{8}~,\qquad \delta=-\frac{N}{2}~. \label{e8thooft}
\end{equation}
Substituting into~\eqref{ais2} then leads to
\begin{equation}
a(\CE _N)=\frac{64}{7}N^{3}+\frac{144}{7}N^{2}+\frac{99}{7}N~.\label{aE8}
\end{equation}

This theory has a (partial) tensor branch that corresponds to separating a single M5-brane from the wall, as illustrated in Figure~\ref{figE8}(b). The degrees of freedom that remain in the deep IR consist of~$N-1$ small~$E_{8}$ instantons~$\CE_{N-1}$ embedded in the wall, together with a free~$(2,0)$ tensor multiplet (for which~$a = 1$) that describes the motion of the separated M5 brane. As was emphasized in~\cite{Intriligator:2014eaa}, the mismatch between the UV and IR 't Hooft anomalies is a perfect square (see appendix~\ref{app} for a discussion of the pre-factor~$\half$),
\begin{equation}
\Delta \CI_8=\frac{1}{2} \left(-Nc_2(R)+\frac{1}{4}p_1(T)\right)^2~.
\end{equation}
Comparing with~\eqref{abcxy}, we find that in our normalization
\begin{equation}
x= -{1\over 8\pi \sqrt{2\pi}} \, N, \qquad y=\frac{1}{4} \cdot {1\over 8\pi \sqrt{2\pi}}~,\label{E8xy}
\end{equation}
where the overall sign of~$x$ and~$y$ has been choose so that the coefficient~$b=\frac{1}{2}(y-x)$ in~\eqref{bxyeq} is positive, as required by unitarity. We can either use~\eqref{aE8}, or~\eqref{deltaaxy} and~\eqref{E8xy}, to compute the change in the~$a$-anomaly along this tensor branch RG flow,
\begin{equation}
\Delta a = a(\CE _N)-a(\CE _{N-1})-1=\frac{24576 \pi ^3}{7} (y-x)^2= \frac{12}{7}(4N+1)^{2}~.
\end{equation}
In accord with the general results of section~\ref{sec:athm}, we see that~$\Delta a$ is positive and proportional to a perfect square. 

It is interesting to investigate RG flows of~$\CE_N$ onto its Higgs branch. Our general results do not automatically imply the~$a$-theorem for such flows, even though we expect it to hold. As was reviewed in section~\ref{sec:moduli}, only the coefficients~$\alpha$ and~$\beta$ in the anomaly polynomial~\eqref{anomalyI} can change along Higgs branch flows, since they are matched by the~$R$-symmetry NG bosons, while~$\gamma$ and~$\delta$ must remain inert. Recalling the discussion around~\eqref{dimhiggs}, we see that the gravitational anomalies of~$\CE_N$ in~\eqref{e8thooft} are consistent with a Higgs branch of dimension~$30 N$. Such a branch exists, and corresponds to dissolving the M5-branes inside the wall (see Figure \ref{figE8}(c)). The geometry of the Higgs branch is described by the moduli space of~$N$ $E_{8}$ instantons, which has quaternionic dimension~$30N$. 
The change in the~$a$-anomaly for the flow onto the Higgs branch is readily computing using~\eqref{aE8} and the~$a$-anomaly~$a = {11 \over 210}$ of a free hypermultiplet (see Table~\ref{known}),
\begin{equation}
\Delta a = a(\CE _N)- 30N \cdot \frac{11}{210}=\frac{64}{7}N^{3}+\frac{144}{7}N^{2}+\frac{88}{7}N~.
\end{equation}
This is manifestly positive, thus verifying the~$a$-theorem for this flow. 

\section{Tensor Branches with Vector Multiplets}
\label{sec:vectors}

So far we have intentionally restricted our attention to~$(1,0)$ RG flows that terminate in a genuine SCFT in the deep IR. In particular, we have only discussed theories whose moduli-space effective actions do not contain massless gauge fields. However, such examples are rare. The simple examples of interacting~$(1,0)$ SCFTs originally discussed in~\cite{Seiberg:1996qx} have vector multiplets on the tensor branch, and there is by now a vast landscape of theories that share this feature. (See the introduction for a brief survey with references.) In this section we extend the preceding analysis to such theories. As discussed in section~\ref{sec:poly}, if the low-energy theory on the tensor branch contains Yang-Mills fields with field strength~$f$, the anomaly polynomial~$\CI_8$ cannot contain any irreducible gauge anomaly~$c_4(f)$. Possible reducible gauge anomalies of the form~$c_2^2(f)$ must be cancelled by a GS-like mechanism~\cite{Green:1984bx,Sagnotti:1992qw}, which leads to a GS term that is related to the gauge kinetic terms by supersymmetry  \cite{Bergshoeff:1996qm,Seiberg:1996qx},
\begin{equation}
- i B \wedge c_2(f) + \varphi \Tr(f^2)~\label{phiFF}.
\end{equation}
The fact that the Yang-Mills kinetic terms depend linearly on the dilaton~$\varphi$ is required by the (spontaneously broken) conformal invariance of the theory, as emphasized in~\cite{Seiberg:1996qx}. 

\subsection{Scale and Conformal Invariance on the Tensor Branch}

\label{sec:sconinv}

As was already mentioned in section~\ref{sec:athm}, the coupled tensor-vector theory in~\eqref{phiFF} is conformally invariant, i.e.~it possesses a well-defined traceless stress tensor, due to the~$\varphi$-dependent kinetic terms for the Yang-Mills field~$f$. On the tensor branch, the vev of~$\varphi$ induces a standard kinetic term for the gauge fields, with gauge coupling~$g^{-2}\sim \langle \varphi \rangle$. Since~$g$ has mass dimension~$-1$, the Yang-Mills theory becomes free in the deep IR. However, IR free gauge theories in~$d > 4$ spacetime dimensions are not genuine CFTs, since their stress tensors have a non-zero trace~$T^\mu_\mu \sim \Tr(f^2)$ (see for instance\cite{ElShowk:2011gz}). (By contrast, free gauge fields are conformally invariant in four dimensions.) They are, in a sense, scale invariant (they possess a conserved dilatation charge), although there is no gauge-invariant scale current. In accord with standard terminology, we will refer to such scale-invariant but non-conformal theories as SFTs. (See for instance~\cite{Luty:2012ww,Fortin:2012hn,Dymarsky:2013pqa,Dymarsky:2014zja,Dymarsky:2015jia} and references therein for a recent discussion of such theories.) 

Even though it is believed that the~$(1,0)$ SCFTs we are considering are conformally invariant, the preceding discussion implies that RG flows onto tensor branches with massless gauge fields terminate in an IR theory that is an SFT, but not a genuine CFT. Similar phenomena occur in the deep IR of many supersymmetric RG flows in five dimensions. By contrast, in four dimensions we are not aware of non-trivial RG flows from an interacting CFT in the UV that terminate in an SFT in the IR. In three dimensions, the gauge coupling is relevant, and there are many examples of non-trivial RG flows from an SFT in the UV to a CFT in the IR. 

In light of the preceding discussion, it is natural to search for an extension of the~$a$-theorem to six-dimensional RG flows between a CFT in the UV and an SFT in the IR. In the remainder of this section, we will formulate and investigate such a generalization, focusing on RG flows of~$(1,0)$ SCFTs onto tensor branches with gauge fields. An immediate challenge is that a suitable analogue of the~$a$-anomaly need not obviously exist for all SFTs. In CFTs, there are many equivalent definitions: we can define the~$a$-anomaly through the anomalous trace of the stress tensor~$T_{\mu\nu}$ in a gravitational background, as in~\eqref{Tanomaly}, or in terms of the four-point function of~$T_{\mu\nu}$ in flat space. For~$(1,0)$ SCFTs, we have argued that the~$a$-anomaly may also be expressed in terms of 't Hooft anomalies as in~\eqref{ais}, which we repeat here,
\begin{equation}
a=\frac{16}{7}\left(\alpha-\beta+\gamma\right) + {6 \over 7} \, \delta~.\label{aisrephere}
\end{equation}
By contrast, in an SFT these various candidate definitions may not agree, or even make sense. For instance, in an SFT the stress tensor need not be traceless, while the definition of~$a$ via~\eqref{Tanomaly} assumes that~$T^\mu_\mu$ is a redundant operator, whose flat-space correlation functions are pure contact terms.

For the purposes of our discussion of RG flows in~$(1,0)$ theories, we choose to define the value of~$a$ in supersymmetric SFTs in terms of their 't Hooft anomalies, via~\eqref{aisrephere}. We can use the 't Hooft anomalies in Table~\ref{known} to compute the value of~$a$ for a free theory of~$n_h$ hypermultiplets, $n_t$ tensor multiplets, and~$n_v$ vector multiplets,
\begin{equation}
a=\frac{1}{210}\left(11n_{h}+199n_{t}-251n_{v}\right)~.\label{afrees}
\end{equation}
The~$a$-anomalies for hypermultiplets and tensor multiplets are well defined and were computed in~\cite{Fradkin:1983tg,Bastianelli:2000hi}. According to our definition, the value of~$a$ for a free vector multiplet (which follows from~\eqref{aisrephere} and the 't Hooft anomalies summarized in~Table~\ref{known}) is negative.

Although surprising, this feature has a precedent in four dimensions. A simple four-dimensional SFT with~$\CN=1$ supersymmetry is the theory of a linear multiplet, which is dual to a free chiral multiplet with a shift symmetry. The shift symmetry forces us to assign vanishing~$U(1)_R$ charge to the scalar in the chiral multiplet, so that its fermionic superpartner has~$R$-charge~$-1$. If we attempt to define the value of the~$a$-anomaluy in this theory by extending the relation~\eqref{a4d} between the~$a$-anomaly and the~$U(1)_R$ symmetry anomalies that holds in SCFTs, we find that~$a = -\frac{3}{16}<0$.

Returning to six dimensions, it is tempting to search for an interpretation of the negativity of the $a$-anomaly, e.g.~by embedding the unitary SFT of a free vector multiplet into a non-unitary CFT as in~\cite{ElShowk:2011gz}.\footnote{~A concrete interpretation along similar lines was subsequently proposed in~\cite{Beccaria:2015uta}.} Another puzzling aspect of~\eqref{afrees} is that it assigns a negative value~$\Delta a < 0$ to Higgsing, where a vector multiplet pairs up with a hypermultiplet to become massive, so that~$\Delta n_h = \Delta n_v = 1$. 

Given the above, we would like to emphasize that the~$a$-anomaly of a genuine unitary CFT in six dimensions is expected to be positive, as has been shown in two and four dimensions~\cite{Zamolodchikov:1986gt,Hofman:2008ar}. In six dimensions, the positivity of~$a$ for all unitary CFTs has not yet been established.\footnote{~The generalization of the arguments in~\cite{Hofman:2008ar} to six dimensions only constrains the~$c$-type Weyl anomalies, but not the~$a$-anomaly \cite{deBoer:2009pn} (see also~\cite{Osborn:2015rna}).} 
(See~\cite{Myers:2010tj} for a general discussion of positivity constraints on~$a$ and~$\Delta a$ from holography.) Below, we will consider explicit examples of unitary RG flows from UV CFTs to IR SFTs such that~$a_{\text{IR}} < 0$. However, in those examples the anomaly deficit~$\Delta a$ between the UV and the IR theories is always sufficiently positive to ensure that~$a_{\text{UV}} >0$. We note in passing that the~$a$-theorem only requires that~$a>0$ for theories that can be deformed to a gapped phase. It follows from the results of~\cite{Cordova:2016xhm,Cordova:2016emh} (see also~\cite{Louis:2015mka}) that this is not possible for~$(1,0)$ theories while maintaining supersymmetry. (See~\cite{Nakayama:2015bwa} for a related discussion in four dimensions.)

\subsection{The $a$-Theorem for Tensor Branch Flows with Vector Multiplets}
\label{sec:avec}

We would now like to state and prove an extension of the~$a$-theorem for flows onto tensor branches of~$(1,0)$ SCFTs that contain vector multiplets, so that the IR theory is an SFT. Within the framework established above, it is straightforward to argue that the inequality~$\Delta a >0$ continues to hold for all RG flows of unitary~$(1,0)$ SCFT onto their tensor branch, even in the presence of vector multiplets. This follows straightforwardly from the definition~\eqref{aisrephere} of~$a$ in terms of 't Hooft anomalies, and the fact that the GS anomaly-matching mechanism for these anomalies implies that~$\Delta \alpha - \Delta \beta + \Delta \gamma$ is a sum of squares, and hence positive (see also section~\ref{sec:highrk}). This argument is not affected by the fact that the IR theory is an SFT. 

More physically, the change~$\Delta a$ as defined by the 't Hooft anomalies continues to determine the six-point, six-derivative scattering amplitudes of the dilaton, even if there are vector fields on the tensor branch and the dilaton couples to them via~$\varphi \Tr(f^2)$, as in~\eqref{phiFF}. This is easy to see by generalizing the discussion at the end of section~\ref{sec:athm}. Recall that the results of~\cite{Chen:2015hpa} imply that there is no independent six-point, six-derivative supervertex for the dilaton, so that this amplitude factorizes through lower-point amplitudes. However, the vertex~$\varphi \Tr(f^2)$ cannot contribute to this factorization. Hence, the presence of vector multiplets does not modify the arguments in sections~\ref{sec:athm} and~\ref{sec:relation} that lead to the relation between~$\Delta a$ and the changes of 't Hooft anomalies in~\eqref{deltafinal}.

\subsection{Unitarity Constraints on Tensor Branch Effective Actions}
\label{sec:aposcons}

As discussed at the end of section~\ref{sec:sconinv}, we expect that~$a > 0$ for unitary CFTs. In the context of RG flows of~$(1,0)$ theories onto their tensor branch, this amounts to 
\begin{equation}
a_{\text{UV}}=a_{\text{IR}}+\Delta a=\frac{16}{7}\left(\Delta \alpha-\Delta \beta +\Delta \gamma \right) + a_{\text{IR}} >0~.  \label{matterrestrict}
\end{equation}
Even though we have argued that~$\Delta a > 0$, the value of~$a_{\text{IR}}$ may be negative if the IR theory contains sufficiently many vector multiplets. In this case \eqref{matterrestrict} constitutes a non-trivial constraint on the matter content and the GS couplings of the tensor-branch effective theory. An example will be discussed below.

\subsection{Example: The Theory of M5-Branes Probing an Orbifold}
\label{sec:orbifold}

An SCFT~$\CT _{N,\Gamma}$ with vector multiplets on the tensor branch can be constructed by placing~$N$ M5-branes on the orbifold singularity $\mathbb{C}^{2}/\Gamma$, where~$\Gamma$ is a discrete subgroup of $SU(2)$.    We define these theories with their free center of mass mode, which is a $(1,0)$ tensor multiplet, included.  Various aspects of these systems have been described in \cite{Blum:1997mm,Brunner:1997gk,Brunner:1997gf,Hanany:1997gh,Heckman:2013pva,DelZotto:2014hpa}.  

The 't Hooft anomalies for these theories were computed in \cite{Ohmori:2014kda}
\begin{align}
& \alpha = |\Gamma|^{2}N^{3}-2N|\Gamma|(r_{\Gamma}+1)+2N+d_{\Gamma}~, \qquad  \beta = N-\frac{1}{2}N|\Gamma|(r_{\Gamma}+1)+\frac{d_{\Gamma}}{2}~,\cr
& \gamma= \frac{1}{8}N+\frac{7d_{\Gamma}}{240}~, \qquad \delta =-\frac{1}{2}N-\frac{d_{\Gamma}}{60}~.
\end{align}
Here $|\Gamma |$ is the order of the discrete group, while~$r_{\Gamma}$ and~$d_\Gamma$ are the rank and dimension of the associated ADE Lie group~$G$ (see Table \ref{tab:GammaGroup} below).
\renewcommand{\arraystretch}{1.5}
\begin{table}[H]
  \centering
  \begin{tabular}{ |c | c | c|c|c|c| }
\hline
$\Gamma$ & $\mathbb{Z}_{k}$& $D_{k}$ & $\mathbb{T}$ & $\mathbb{O}$ &  $\mathbb{D}$\\
\hline
$G$ &$SU(k)$ & $SO(2k+2)$ & $E_{6}$ & $E_{7}$ & $E_{8}$ \\
\hline
$|\Gamma|$ &$k$ &$4k$&$24$&$48$&$120$\\
\hline
$r_{\Gamma}$ & $k-1$& $k+1$& $6$&$7$&$8$ \\
\hline
$d_{\Gamma}$ & $k^{2}-1$& $2k^{2}+3k+1$& $78$&$133$&$248$ \\
\hline
\end{tabular}
  \caption{Group theory coefficients for discrete groups $\Gamma.$  The group $D_{k}$ is the binary dihedral group of order $4k$.  The groups $\mathbb{T}, \mathbb{O}, \mathbb{D}$ are respectively the binary tetrahedral, octahedral, and dodecahedral subgroups of $SU(2)$.  The ADE group $G$ is associated to $\Gamma$ by the McKay correspondence. }
  \label{tab:GammaGroup}
\end{table}
The formula \eqref{ais} then determines the Weyl $a$-anomaly:
\begin{equation}
a(\mathcal{T}_{N,\Gamma})=\frac{16}{7}N^{3}|\Gamma|^{2}-\frac{24}{7}N|\Gamma|(r_{\Gamma}+1)+\frac{15}{7}N+\frac{251}{210}d_{\Gamma}~.\label{aTNGamma}
\end{equation}
When~$\Gamma$ is trivial, this reduces to the $(2,0)$ conformal anomaly \cite{Cordova:2015vwa}.  The fact that $a_{\CT_N}\approx \frac{16}{7}N^3$ for these theories was first found via AdS/CFT~\cite{Henningson:1998gx,Bastianelli:2000hi}. The extra factor of~$|\Gamma|^2$ that multiplies the leading~$N^3$ behavior in the orbifold case can similarly be understood from holography on $AdS_7\times (S^4/\Gamma)$: the modified volume of $S^4/\Gamma$ contributes a factor of~$1/|\Gamma|$, but we must also change~$N\to |\Gamma|N$ to get the same amount of flux (i.e.\,M5-branes). Note that, as expected $a(\mathcal{T}_{N,\Gamma})>0$ in all cases, consistent with the interpretation of this system as a unitary SCFT.

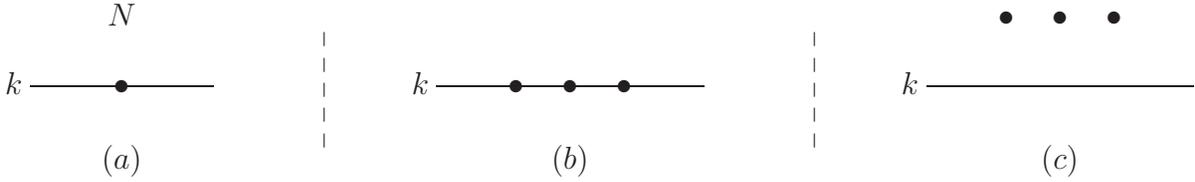
\begin{figure}[h]
\[
\xymatrix  @R=1pc {
& {N} && {\phantom{\bullet}}\ar@{--}[dd] & & && {\phantom{\bullet}}\ar@{--}[dd] &  & {\bullet~~~ \bullet~~~ \bullet} &  \\
{k}\ar@{-}[rr] &{\bullet}&{\phantom{\bullet}}&& {k}\ar@{-}[rr] &{\bullet~~~ \bullet~~~ \bullet}&{\phantom{\bullet}}&&   {k}\ar@{-}[rr] &&{\phantom{\bullet}}\\
& {(a)}& & {\phantom{\bullet}} && {(b)}& & {\phantom{\bullet}} && {(c)}&
}  \]
 \caption{The orbifold theory and its RG flows.  In (a) there are $N$ NS5-branes (represented by dots) embedded in a stack of $k$ D6-branes.  In (b) the flow onto the tensor branch is obtained by separating the NS5-branes inside the D6-branes.  In (c) the flow onto a mixed branch is obtained by moving the NS5-branes off of the D6-branes. }
  \label{figORB}
 \end{figure}

To study various RG flows, let us restrict to the special case $\Gamma=\mathbb{Z}_{k}$.  By reducing from M-theory to type IIA, this theory can then be given a brane interpretation as $N$ coincident NS5-branes embedded inside a stack of $k$ D6-branes (see e.g. Figure \ref{figORB}(a).)

The low-energy field content of the tensor branch of this theory can be readily seen from the brane diagram illustrated in Figure \ref{figORB}(b).  By separating the stack of M5-branes along the $\mathbb{C}^{2}/\mathbb{Z}_{k}$ singularity we arrive at a theory with the following fields:
\begin{itemize}
\item $N$ (1,0) tensor multiplets.  The expectation values of the scalars in these multiplets parameterize the motion of the NS5-branes along the D6 branes.
\item Vector multiplets for $(N-1)$ copies of $SU(k)$ gauge groups.  These arise on the finite slabs of D6-branes bounded by NS5-branes.  The non-compact D6-branes give rise to an $SU(k)\times SU(k)$ global symmetry.  Note that all $U(1)$ gauge and global symmetries of this system are anomalous and lifted from the spectrum.
\item $Nk^{2}$ hypermultiplets.   These arise from string modes connecting adjacent slabs of D6-branes which are separated by an NS5-brane.  They therefore transform as bifundamentals under adjacent $SU(k)$ groups.
\end{itemize}
From this description of the matter content, we may readily evaluate the infrared value of the $a$-anomaly on the tensor branch as
\begin{equation}
a_\text{Tensor}=\frac{15}{7}N-\frac{8}{7}Nk^{2}+\frac{251}{210}(k^{2}-1)~.
\end{equation}
For general values of $N$ and $k>1$, this expression can be negative, due to the negative contribution of the vector multiplets. However, the change~$\Delta a$, which can be computed using \eqref{aTNGamma}, is positive, as required by our general arguments. 

The change in the 't Hooft anomalies between the UV theory at the origin and the IR theory on the tensor branch is 
\begin{equation}
\Delta \alpha=Nk^{2}(N^{2}-1)~, \hspace{.5in} \Delta \beta=0~, \hspace{.5in} \Delta \gamma =0~.
\end{equation}
The constraint of section~\ref{sec:aposcons} then reads 
\begin{equation}
a_\text{UV}=\frac{16}{7}\Delta \alpha +a_\text{Tensor}>0~,
\end{equation}
which is indeed satisfied, even though~$a_\text{Tensor}$ can be negative.

We can also flow onto a mixed branch.  This is achieved by a motion of the of the NS5-branes transverse to the singularity as illustrated in Figure \ref{figORB}(c).  At low-energies, the resulting theory is described by $N$ free $(1,0)$ tensor multiplets together with $N$ hypermultiplets probing the singularity $\mathbb{C}^{2}/\mathbb{Z}_{k}.$ We therefore have
\begin{equation}
a_\text{Mixed}=N~,
\end{equation}
which is positive, since the low-energy theory does not contain any vector multiplets. Note that the naive extension of the~$a$-theorem to flows from the tensor to the mixed branche is false because~$a_\text{Tensor}< a_\text{Mixed}$. However, as expected, the full RG flow from the UV CFT~$\mathcal{T}_{N,\Gamma}$ onto the mixed branch in the IR does satisfy the $a$-theorem~$a(\mathcal{T}_{N,\Gamma})>a_\text{Mixed}$.  

It would be interesting to consider RG flows of other theories onto Higgs and mixed branches (see e.g. \cite{Gaiotto:2014lca,Heckman:2015ola}), and further investigate the monotonicity properties of $a$.

\section*{Acknowledgements}
 
\noindent We are grateful to J.~Maldacena and A.~Zhiboedov for discussions.  We would also like to thank~J.~Heckman and C.~Herzog for sharing a draft on related topics.  The work of CC is supported by a Junior Fellowship at the Harvard Society of Fellows.  TD is supported by the Fundamental Laws Initiative of the Center for the Fundamental Laws of Nature at Harvard University, as well as DOE grant DE-SC0007870 and NSF grants PHY-0847457, PHY-1067976, and PHY-1205550. The work of KI is supported in part by the US Department of Energy under UCSDs contract de-sc0009919.

\appendix

\section{Green-Schwarz Mechanism for Chiral Scalars}

\label{app}

Throughout the paper, various Green-Schwarz (GS) terms for chiral two-form gauge fields, with self-dual three-form field strengths, in six spacetime dimensions have played an important role. Their contribution to the anomaly eight-form polynomial was explained in section~\ref{sec:moduli}, including a crucial factor of~$\half$ in~\eqref{dixsq} that follows from the self-duality constraint. Here we briefly review an analogous phenomenon in the simpler context of chiral scalars in two spacetime dimensions. It may be helpful to keep in mind the description of the chiral boson in terms of a free chiral fermion. Here we emphasize the bosonic point of view, because of the analogy with chiral two-form gauge fields in six dimensions. 

It is convenient to describe the chiral scalar using a Lagrangian, at the expense of manifest Lorentz invariance. Following~\cite{Floreanini:1987as,Belov:2006jd} (see also~\cite{moorefklect}), we consider the following Lagrangian for a real field~$\phi(x,y)$ (its relation to the chiral scalar will be described below),
\begin{equation}\label{philag}
\SL = {\Omega \over 2} \, \d_x\phi \left(\d_x \phi +  i \sigma \d_y \phi\right)~, \qquad \sigma = \pm ~.
\end{equation}
We work in Euclidean signature.\footnote{~Wick rotating to Lorentzian signature replaces~$\d_y \rightarrow -i \d_t$, so that the Lagrangian~\eqref{philag} becomes real.} The normalization factor~$\Omega > 0$ is analogous to the matrix~$\Omega_{IJ}$ that appears in~\eqref{SBwedgeX} and~\eqref{dixsq}. (Canonically normalized kinetic terms are obtained by setting~$\Omega = 1$.) The sign factor~$\sigma$ will turn out to determine the chirality of the scalar. This can be seen by varying~\eqref{philag} to obtain the following equation of motion,
\begin{equation}\label{eom}
\left(\d_x +  i  \sigma \d_y\right) \d_x \phi = 0~.
\end{equation}
We can therefore introduce a one-form~$J$ that satisfies
\begin{equation}\label{jdef}
J = \, \d_x \phi \left(dx +  i \sigma dy\right)~, \qquad * J = -i \sigma J~, \qquad d J = 0~.
\end{equation}
The (anti-) self-dual one-form~$J$ is simply the conserved current corresponding to the~$U(1)$ flavor symmetry carried by the chiral scalar. 

It is a standard fact that the chiral~$U(1)$ current in~\eqref{jdef} has a non-zero 't Hooft anomaly, which can be exhibited by coupling~$J$ to a background gauge field~$A$, 
\begin{equation}\label{sourceS}
\SL_{A} = \SL + \Omega \left( J_x A_x + J_y A_y\right) =  {\Omega \over 2} \, \d_x\phi \left(\d_x \phi +  i \sigma \d_y \phi\right) +  \Omega \, \d_x \phi \left(A_x + i \sigma A_y\right)~.
\end{equation}
Here we have included a factor of~$\Omega$ in the~$J$-$A$ coupling to mirror the conventions used in~\eqref{SBwedgeX}. The equations of motion~\eqref{eom} are deformed to
\begin{equation}\label{eomii} \d_x \Big(\left(  \d_x \phi +  A_x\right) + i \sigma \left( \d_y \phi +  A_y\right) \Big) = 0~.
\end{equation}
They are manifestly invariant under background gauge transformations
\begin{equation} 
\delta A = d \lambda~, \qquad 
\delta \phi = - \lambda~, \qquad \lambda = \lambda(x,y)~.
\end{equation} 
By contrast, the Lagrangian~\eqref{sourceS} is not invariant,
\begin{equation}\label{Svar}
\delta \SL_A = - \Omega\,  \d_x \lambda \left(A_x + i \sigma A_y\right) + \left(\text{total derivative}\right)~.
\end{equation}
As befits an anomaly, this non-invariance cannot be removed using the available local counterterms~$A_x^2, A_y^2$, and~$A_x A_y$. However, they can be tuned to covariantize the variation~\eqref{Svar}, 
\begin{equation}\label{finaldeltas}
\delta \left(\SL_A + {\Omega \over 2 } A_x^2 + {i \sigma \Omega \over 2} A_x A_y\right) = {i  \sigma  \over 2}\, \Omega \,  \lambda \, F + \left(\text{total derivative}\right)~, \qquad F = dA~.
\end{equation}
This expression manifests the factor of~$\half$ discussed at the beginning of this appendix. Note that the sign of the anomaly depends on the chirality~$\sigma$ of the scalar, consistent with the fact that the shift symmetry of a non-chiral scalar does not have an anomaly. The variation~\eqref{finaldeltas} can be accounted for by a contribution to the anomaly four-form polynomial. In the conventions of section~\ref{sec:poly},
\begin{equation}
\Delta \CI_4 = {\sigma \over 2} \cdot {\Omega \over 2 \pi} \, F \wedge F~.
\end{equation}
This equation is the two-dimensional analogue of the six-dimensional formula~\eqref{dixsq}. There the overall sign was also fixed by the chirality of the two-form gauge fields in tensor multiplets, which is determined by supersymmetry.

\bibliographystyle{utphys}
\bibliography{atheorem10}

\end{document}